\documentclass[letterpaper]{article} % DO NOT CHANGE THIS
\usepackage{aaai24}  % DO NOT CHANGE THIS 
\usepackage{times}  % DO NOT CHANGE THIS
\usepackage{helvet}  % DO NOT CHANGE THIS
\usepackage{courier}  % DO NOT CHANGE THIS
\usepackage[hyphens]{url}  % DO NOT CHANGE THIS
\usepackage{graphicx} % DO NOT CHANGE THIS
\usepackage{natbib}  % DO NOT CHANGE THIS AND DO NOT ADD ANY OPTIONS TO IT
\usepackage{caption} % DO NOT CHANGE THIS AND DO NOT ADD ANY OPTIONS TO IT
\usepackage{algorithm}
\usepackage[noend]{algorithmic}

\usepackage{newfloat}
\usepackage{listings}
\DeclareCaptionStyle{ruled}{labelfont=normalfont,labelsep=colon,strut=off} % DO NOT CHANGE THIS
\floatstyle{ruled}
\newfloat{listing}{tb}{lst}{}
\floatname{listing}{Listing}
\usepackage{multirow}
\usepackage{xspace}
\usepackage{verbatimbox}
\usepackage[table,xcdraw]{xcolor}
\usepackage{cleveref}
\usepackage{etoolbox}
\usepackage{subcaption}
\usepackage{anyfontsize}
\usepackage{siunitx}
\usepackage{diagbox}
\newcommand{\nomos}{\textsc{Nomos}\xspace}
\newcommand{\tcbase}{\textsc{TransCoder}\xspace}
\newcommand{\tcir}{\textsc{TransCoder-IR}\xspace}
\newcommand{\dobf}{\textsc{DOBF}\xspace}
\newcommand{\starcoder}{\textsc{StarCoder}\xspace}
\definecolor{gray}{rgb}{0.5, 0.5, 0.5}
\definecolor{light-gray}{gray}{0.77}
\definecolor{BrickRed}{rgb}{0.8, 0.25, 0.33}
\definecolor{Black}{rgb}{0.0, 0.0, 0.0}
\definecolor{DarkBlue}{rgb}{0.0, 0.0, 0.55}
\definecolor{Crimson}{rgb}{0.86, 0.08, 0.24}
\definecolor{SlateGrey}{rgb}{0.44, 0.5, 0.56}
\definecolor{lightorange}{HTML}{FFB74D}
\definecolor{blue}{rgb}{0.0, 0.0, 1.0}
\definecolor{magenta}{rgb}{0.79, 0.08, 0.48}
\definecolor{darkgreen}{rgb}{0.0, 0.2, 0.13}
\definecolor{darkraspberry}{rgb}{0.53, 0.15, 0.34}
\definecolor{antiquefuchsia}{rgb}{0.57, 0.36, 0.51}
\usepackage{listings}
\lstdefinestyle{basic}{%
  keywordstyle     = [2]\color{teal}\bfseries,%
  keywordstyle     = [3]\color{BrickRed}\bfseries,%
  keywordstyle     = [4]\color{darkraspberry}\bfseries,%
  keywordstyle     = \bfseries\color{DarkBlue},%
  commentstyle     = \ttfamily\color{Black!50}\lst@ifdisplaystyle\scriptsize\fi,%
  basicstyle       = \small\ttfamily\lst@ifdisplaystyle\scriptsize\fi,%
  emph             = {int,char,double,float,unsigned,void,bool},%
  emphstyle        = {\color{teal}\bfseries},%
  stringstyle      = \color{antiquefuchsia}\itshape,%
  columns          = [c]fixed,%
  aboveskip        = 0mm,%
  belowskip        = 2mm,%
  keepspaces       = true,%
  mathescape       = true,%
  escapechar       = @,%
  tabsize          = 2,%
  numbers          = left,%
  numberstyle      = \tiny\color{Black!50},%
  numbersep        = 4pt,%
  stepnumber       = 1,%
  firstnumber      = 1,%
  showstringspaces = false,%
  captionpos       = b,%
  extendedchars    = true,%
  upquote          = true,%
  abovecaptionskip = 0mm,%
  belowcaptionskip = 0mm,%
  moredelim        = **[is][{\btHL[fill=light-gray]}]{Â°}{Â°},%
}

\lstdefinestyle{nomos}{%
  language         = Python,%
  style            = basic,%
  morekeywords     = [1]{var,input,output,ensures,import,requires},%
  morekeywords     = [2]{randInt,getFeat,label,strConcat,setFeat,blur,wNoise,relax,unrelax,renameParam,renameParam,addParam,addConditional,chBranchCond,addLoop,rmLoop,merge,arity,numLoops,numConditionals,compiles,run,retValues},%
  morekeywords     = [3]{MAX_INT,null},%
  morekeywords     = [4]{play,predict,transpile},%
  xleftmargin=1.2em,%
}

\lstdefinestyle{basic-inline}{%
  keywordstyle     = [2]\bfseries,%
  keywordstyle     = [3]\bfseries,%
  keywordstyle     = [4]\bfseries,%
  keywordstyle     = \bfseries,%
  commentstyle     = \ttfamily\lst@ifdisplaystyle\scriptsize\fi,%
  basicstyle       = \small\ttfamily\lst@ifdisplaystyle\scriptsize\fi,%
  emph             = {int,char,double,float,unsigned,void,bool},%
  emphstyle        = {\bfseries},%
  stringstyle      = \itshape,%
  columns          = [c]fixed,%
  aboveskip        = 0mm,%
  belowskip        = 2mm,%
  keepspaces       = true,%
  mathescape       = true,%
  escapechar       = @,%
  tabsize          = 2,%
  numbers          = left,%
  numberstyle      = \tiny\color{Black!50},%
  numbersep        = 4pt,%
  stepnumber       = 1,%
  firstnumber      = 1,%
  showstringspaces = false,%
  captionpos       = b,%
  extendedchars    = true,%
  upquote          = true,%
  abovecaptionskip = 0mm,%
  belowcaptionskip = 0mm,%
  moredelim        = **[is][{\btHL[fill=light-gray]}]{Â°}{Â°},%
}

\lstdefinestyle{nomos-inline}{%
  language         = Python,%
  style            = basic-inline,%
  morekeywords     = [1]{var,input,output,ensures,import,requires},%
  morekeywords     = [2]{randInt,getFeat,label,strConcat,setFeat,blur,wNoise,relax,unrelax,renameParam,renameParam,addParam,addConditional,chBranchCond,addLoop,rmLoop,merge,arity,numLoops,numConditionals,compiles,run,retValues},%
  morekeywords     = [3]{MAX_INT,null},%
  morekeywords     = [4]{play,predict,transpile},%
  xleftmargin=1.2em,%
}

\newcommand\code[1]{\lstinline[style=nomos-inline]{#1}\xspace}

\lstdefinestyle{python}{%
  language         = Python,%
  style            = basic,%
  xleftmargin=1.2em,%
}
\newtoggle{longversion}
\settoggle{longversion}{true}

\newcommand{\app}[1]{\iftoggle{longversion}{
\Cref{#1}%
}
{
\cite{EniserWuestholz2023}%
}}

\newcommand{\apps}[2]{\iftoggle{longversion}{
Appendices~\ref{#1} and \ref{#2}%
}
{
\cite{EniserWuestholz2023}%
}}

\title{Automatically Testing Functional Properties of Code Translation Models}
\author{
    Hasan Ferit Eniser\textsuperscript{\rm 1}, Valentin W{\"u}stholz\textsuperscript{\rm 2}, Maria Christakis\textsuperscript{\rm 3}
}
\affiliations{
    \textsuperscript{\rm 1} MPI-SWS ,Germany \\
    \textsuperscript{\rm 2} ConsenSys, Austria \\
    \textsuperscript{\rm 3} TU Wien, Austria \\
    hfeniser@mpi-sws.org, valentin.wustholz@consensys.net, maria.christakis@tuwien.ac.at
}

\usepackage{bibentry}
\begin{document}

\maketitle

%%----------------------------------------------------------------------------
\begin{abstract}
    Large language models are becoming increasingly practical for translating code across programming languages, a process known as \emph{transpiling}. Even though automated transpilation significantly boosts developer productivity, a key concern is whether the generated code is correct. Existing work initially used manually crafted test suites to test the translations of a small corpus of programs; these test suites were later automated. In contrast, we devise the first approach for automated, functional, property-based testing of code translation models. Our general, user-provided specifications about the transpiled code capture a range of properties, from purely syntactic to purely semantic ones. As shown by our experiments, this approach is very effective in detecting property violations in popular code translation models, and therefore, in evaluating model quality with respect to given properties. We also go a step further and explore the usage scenario where a user simply aims to obtain a correct translation of some code with respect to certain properties without necessarily being concerned about the overall quality of the model. To this purpose, we develop the first property-guided search procedure for code translation models, where a model is repeatedly queried with slightly different parameters to produce alternative and potentially more correct translations. Our results show that this search procedure helps to obtain significantly better code translations.
\end{abstract}
%%----------------------------------------------------------------------------

%%----------------------------------------------------------------------------
\section{Introduction}
\label{sect:intro}
%%----------------------------------------------------------------------------

Large language models (LLMs) are becoming highly relevant for translating code across programming languages, a process also known as \emph{transpiling}. Transpilation is typically used to translate an existing software system written in an obsolete programming language into a modern language or to integrate code bases written in different languages into one. On one hand, automated transpilation tremendously increases developer productivity. On the other hand, a key concern is whether the transpiled code is correct.

Existing work used manually crafted test suites~\cite{RoziereLachaux2020} to assess the quality of translations for individual functions.
These test suites were later generated automatically using search-based testing~\cite{RoziereZhang2022}. However, given the small corpus of functions, it remains unclear how well the models generalize to other code.

\textbf{Our approach.} In this paper, we automatically test functional properties of code translation models themselves. To this end, we extend \nomos~\cite{ChristakisEniser2023}, an open-source framework for expressing functional-correctness properties of machine learning models and automatically testing models against these properties. In particular, \nomos uses a declarative, domain-agnostic specification language for writing \emph{hyperproperties}~\cite{ClarksonSchneider2008} (or \emph{$k$-safety properties}), which capture functional correctness by reasoning about $k$ model executions. As an example, consider a recidivism-risk model predicting whether a criminal is likely to re-offend. The property that ``if a criminal's number of priors increases, then their recidivism risk should not decrease" is a \emph{2-safety property}---we need two model executions to detect a violation of this property, both of which take as input the same criminal but one with an increased number of priors. \nomos has been used to effectively test models from various application domains, namely action policies as well as models that take as input tabular data, images, speech, and natural language.

Here, we build on \nomos to enable it to express and validate a wide range of $k$-safety properties about code translation models, ranging from purely \emph{syntactic} to purely \emph{semantic} properties of the transpiled code.
An example of a purely syntactic property is that ``the number of loops in the translated code should match the number of loops in the original code"; a purely semantic property is that ``the translated code should produce the same return values as the original code for a given set of inputs". We can also express \emph{compilation preservation}, that is, ``if the original code compiles, the translated code should also compile". Note that compilers typically perform both syntactic (e.g., parsing) and semantic (e.g., type checking) analyses, and thus, compiler preservation could be viewed as a hybrid property, between purely syntactic and purely semantic ones. These are examples of standard (1-)safety properties---they can be validated by a single model execution that generates the translated code.

A $k$ greater than $1$ enables specifying the expected model behavior more comprehensively. For instance, violations of the above compilation property may be unavoidable for some programs---e.g., the source program uses a library function for which there is no comparable version in the target language. Writing $k$-safety properties allows not labeling such unavoidable model behavior as ``buggy" and identifying more severe issues. An example of such a 2-safety property is: ``given a program $P$, if a function parameter is renamed in $P'$ and compilation preservation holds for one of the two programs, then it should also hold for the other". More specifically, $P'$ is an equivalent variant of $P$ that is randomly generated by renaming a function parameter in $P$. A property violation is detected when $P'$ (conversely, $P$) and its translated version compile whereas the translated version of $P$ (conversely, $P'$) does not compile. Such a violation indicates more severe buggy behavior of the model than if, say, $P$ compiled but its translated version did not (1-safety). After all, we know that compilation preservation must hold when applying a simple, equivalent transformation.

Properties like these can be succinctly expressed and validated in our \nomos extension. On a high level, the user provides a code translation model and the ($k$-safety) properties that it should satisfy. The output is a set of tests that violate the given properties. For example, for the above 2-safety property, program $P$ would be selected by \nomos from an existing corpus, such as the model test set, and $P'$ would be automatically generated. As output, \nomos would produce tests, each comprising 2 inputs to the model, namely $P$ and $P'$, for which the property fails. Under the hood, we automatically translate the given properties into a \emph{test harness}, that is, code that uses \emph{metamorphic testing}~\cite{ChenCheung1998,SeguraFraser2016} to generate inputs to the model under test and validate its outputs against an oracle expressing the expected behavior.

As our results show, our approach is very effective in detecting property violations in popular code translation models, such as \tcbase~\cite{RoziereLachaux2020}, \dobf~\cite{LachauxRoziere2021}, \tcir~\cite{SzafraniecRoziere2023}, and \starcoder~\cite{LiAllal2023}. When testing these models against 38 properties that we specified, we detect thousands of violations. When used in this way, our approach can therefore help evaluate the quality of the models under test with respect to given properties. Any detected violations could even help repair the models although it could be costly~\cite{OuyangWu2022}.

In this paper, we explore another usage scenario in which the user aims to obtain a correct translation of some code with respect to a set of properties without updating the model. As a result, we devise a property-guided search procedure for code translation models, where a model is repeatedly queried with slightly different parameters (e.g., temperature) to produce an alternative translation that potentially satisfies the desired properties. Note that, for this scenario, there are certain guidelines for writing $k$-safety properties that are compatible with our search (see \Cref{sect:method}).

In summary, we make the following contributions:
\begin{itemize}
    \item We devise the first approach for automated, functional, property-based testing of code translation models, which we implement as an extension of \nomos. Our implementation is publicly available\footnote{\url{https://github.com/Rigorous-Software-Engineering/nomos}}.
    \item We present the first formalization of $k$-safety properties for this domain, ranging from purely syntactic to purely semantic ones.
    \item We develop the first property-guided search procedure for code translation models to generate alternative translations that potentially satisfy a given set of properties.
    \item We evaluate the effectiveness of our approach in detecting violations of 38 properties across four state-of-the-art code translation models. We also show that our search procedure can help obtain significantly better translations for a given user-provided program.
\end{itemize}

%%----------------------------------------------------------------------------
\section{Related Work}
%%----------------------------------------------------------------------------

\begin{figure*}[t!]
\begin{subfigure}[b]{.54\textwidth}
\begin{lstlisting}[style=nomos]
input pj; @ \label{line:syn1-1} @
output pc; @ \label{line:syn1-2} @
{ @ \label{line:syn1-3} @
  pc = transpile(pj, "java", "cpp") @ \label{line:syn1-4} @
} @ \label{line:syn1-5} @
ensures numConditionals(pj, "java") @ \label{line:syn1-6} @
  == numConditionals(pc, "cpp"); @ \label{line:syn1-7} @
\end{lstlisting}
\caption{Syntactic 1-safety property.}
\label{fig:syntactic1}
\end{subfigure}%
\begin{subfigure}[b]{.45\textwidth}
\begin{lstlisting}[style=nomos]
input pj; @ \label{line:sem1-1} @
requires compiles(pj, "java"); @ \label{line:sem1-2} @
output pc; @ \label{line:sem1-3} @
{ @ \label{line:sem1-4} @
  pc = transpile(pj, "java", "cpp") @ \label{line:sem1-5} @
} @ \label{line:sem1-6} @
ensures compiles(pc, "cpp") @ \label{line:sem1-7} @
  ==> retValues(pj, "java") == retValues(pc, "cpp");
\end{lstlisting}
\caption{Semantic 1-safety property.}
\label{fig:semantic1}
\end{subfigure}\\\\
\begin{subfigure}[b]{.54\textwidth}
\begin{lstlisting}[style=nomos]
input pj1; @ \label{line:syn2-1} @
var pj2 := addConditional(pj1, "java"); @ \label{line:syn2-2} @
output pp1; @ \label{line:syn2-3} @
output pp2; @ \label{line:syn2-4} @
{ @ \label{line:syn2-5} @
  pp1 = transpile(pj1, "java", "py") @ \label{line:syn2-6} @
  pp2 = transpile(pj2, "java", "py") @ \label{line:syn2-7} @
} @ \label{line:syn2-8} @
ensures numLoops(pp1, "py") == numLoops(pp2, "py"); @ \label{line:syn2-9} @
\end{lstlisting}
\caption{Syntactic 2-safety property.}
\label{fig:syntactic2}
\end{subfigure}%
\begin{subfigure}[b]{.45\textwidth}
\begin{lstlisting}[style=nomos]
input pj1; @ \label{line:sem2-1} @
var pj2 := renameParam(pj1, "java"); @ \label{line:sem2-2} @
requires pj2 != null; @ \label{line:sem2-3} @
output pp1; @ \label{line:sem2-4} @
output pp2; @ \label{line:sem2-5} @
{ @ \label{line:sem2-6} @
  pp1 = transpile(pj1, "java", "py") @ \label{line:sem2-7} @
  pp2 = transpile(pj2, "java", "py") @ \label{line:sem2-8} @
} @ \label{line:sem2-9} @
ensures compiles(pp1, "py") && compiles(pp2, "py") @ \label{line:sem2-10} @
  ==> retValues(pp1, "py") == retValues(pp2, "py");
\end{lstlisting}
\caption{Semantic 2-safety property.}
\label{fig:semantic2}
\end{subfigure}
\caption{Example $k$-safety specifications for code translation models.}
\label{fig:examples}
\end{figure*}

% In this section, we focus on related work that aims to address correctness concerns when using LLMs for automated code translation or generation.

\noindent
\textbf{Code translation.} The most closely related work uses LLMs for \emph{code translation}~\cite{RoziereLachaux2020, LachauxRoziere2021, RoziereZhang2022,SzafraniecRoziere2023} and manually written or automatically generated tests~\cite{RoziereZhang2022} to evaluate semantic correctness of the translated code (similar to our semantic 1-safety property) for a relatively small, curated corpus of programs. In contrast, we focus on testing the correctness of the models themselves. Specifically, we enable \nomos to express a much broader and more comprehensive set of correctness properties; we also automatically generate new programs, instead of only relying on an existing, curated corpus.
%
% RoziereLachaux2020: https://proceedings.neurips.cc/paper/2020/file/ed23fbf18c2cd35f8c7f8de44f85c08d-Paper.pdf
% LachauxRoziere2021: https://openreview.net/pdf?id=3ez9BSHTNT
% RoziereZhang2022: https://openreview.net/pdf?id=cmt-6KtR4c4
% SzafraniecRoziere2023: https://openreview.net/pdf?id=XomEU3eNeSQ

\textbf{Code generation.} There is also work based on LLMs for \emph{code generation} that uses prompts in (primarily) natural language and evaluates the correctness of the generated code. \textsc{HumanEval}~\cite{ChenTworek2021} is a popular benchmark in this context, but it uses a small number of tests to evaluate the semantic correctness of the generated code (similar to our semantic 1-safety property). \textsc{EvalPlus}~\cite{LiuXia2023} extends \textsc{HumanEval} to obtain more comprehensive benchmarks---it uses fuzzing to automatically generate many more tests. Other work~\cite{CassanoGouwar2023, AthiwaratkunGouda2023} has proposed methods for extending benchmarks, such as \textsc{HumanEval} and MBPP~\cite{AustinOdena2021}, to more programming languages. ReCode~\cite{WangLi2023} checks robustness properties (somewhat similar to our semantic 2-safety properties, but for code generation instead of code translation) by slightly perturbing the prompts through over 30 transformations.
%
% ChenTworek2021: https://arxiv.org/pdf/2107.03374.pdf
% AustinOdena2021: https://arxiv.org/pdf/2108.07732.pdf
% LiuXia2023: https://arxiv.org/pdf/2305.01210.pdf
% CassanoGouwar2023: https://arxiv.org/pdf/2208.08227.pdf
% AthiwaratkunGouda2023: https://arxiv.org/pdf/2210.14868.pdf
% WangLi2023: https://arxiv.org/pdf/2212.10264.pdf

\textbf{Constraining LLM outputs.} Constraining LLM outputs to enforce certain validity criteria has also been explored. For instance, in the context of programming languages, such criteria may enforce syntactic or semantic constraints on the output programs or completions~\cite{ScholakSchucher2021, PoesiaPolozov2022}. On a high level, our search procedure pursues a similar goal in the context of code translation, but phrases it as an optimization problem that aims to minimize the number of violated properties. More generally, such validity criteria can also consist of a grammar~\cite{ShinLin2021} or a domain-specific query language~\cite{BeurerKellnerFischer2023}.
%
% Synchromesh: https://arxiv.org/abs/2201.11227
% Picard: https://arxiv.org/pdf/2109.05093.pdf
% BeurerKellnerFischer2023: https://dl.acm.org/doi/pdf/10.1145/3591300
% ShinLin2021: https://aclanthology.org/2021.emnlp-main.608.pdf
    
% The following seem much less relevant:
% Beyond all these, there exists various prompting methods~\cite{WeiWang2022,ShinLin2021,YaoZhao2023} for getting better predictions, and adversarial attacks to LLMs~\cite{ZouWang2023,MausChao2023,WeiHaghtalab2023}, however, they are not closely related to this study.
%
% WeiWang2022: https://openreview.net/pdf?id=_VjQlMeSB_J
% YaoZhao2023: https://openreview.net/pdf?id=WE_vluYUL-X
% ZouWang2023: https://arxiv.org/pdf/2307.15043.pdf
% MausChao2023: https://arxiv.org/pdf/2302.04237.pdf
% WeiHaghtalab2023: https://arxiv.org/pdf/2307.02483.pdf
% \cite{JonesDragan2023} cast auditing LLMs as a discrete optimization problem and aim to reveal unexpected behaviors such as toxic outputs.
%
% JonesDragan2023: https://arxiv.org/abs/2303.04381

%%----------------------------------------------------------------------------
\section{Approach}
\label{sect:method}
%%----------------------------------------------------------------------------

In this section, we first give an overview of our specifications for code translation models through examples (\Cref{sect:specs}). We then describe our testing (\Cref{sect:checking-procedure}) and search (\Cref{sect:search-procedure}) procedures in detail.

%%----------------------------------------------------------------------------
\subsection{Specifications}
\label{sect:specs}
%%----------------------------------------------------------------------------

We introduce our extended \nomos specification language through four example properties, namely, two 1-safety properties (a syntactic and a semantic one) and two 2-safety properties (again, a syntactic and a semantic one). On a high level, each specification typically consists of:
\begin{itemize}
    \item A \emph{precondition}, which expresses the conditions under which the model under test should be called;
    \item A block of arbitrary source code (written in Python), which invokes the model under test;
    \item A \emph{postcondition}, which expresses the safety property that the model should satisfy.
\end{itemize}
For a formal description of the language prior to our extension, we refer the reader to the \nomos paper~\cite{ChristakisEniser2023}. We describe our
extensions in the next section.

First, consider the syntactic 1-safety property shown in \Cref{fig:syntactic1} expressing that, when transpiling Java code into C++, the number of conditionals in the input (Java) program should match the number of conditionals in the output (C++) program. \Cref{line:syn1-1} declares the input program \code{pj} and \cref{line:syn1-2} the corresponding output program \code{pc}. These declarations are followed by the block of Python code (within curly braces), which invokes the model under test to transpile \code{pj} and assigns the resulting code to \code{pc}---see lines~\ref{line:syn1-3}--\ref{line:syn1-5}. On lines~\ref{line:syn1-6}--\ref{line:syn1-7}, the \code{ensures} clause expresses the postcondition that the number of conditionals in \code{pj} should be equal to the number of conditionals in \code{pc}. Note that there is no precondition in this property, i.e., the precondition is true.

\Cref{fig:semantic1} shows a semantic 1-safety property expressing, with a precondition on \cref{line:sem1-2}, that the model should be called with a compiling Java program. Note that preconditions are specified with \code{requires} clauses. The postcondition says that, if the resulting C++ program is also compiling, then the return values of the two programs should match. In other words, given the same input values, the two programs should return the same output values.

A syntactic 2-safety property about a Java-to-Python translation is shown in \Cref{fig:syntactic2}. On \cref{line:syn2-1}, the property declares an input program \code{pj1}. Unlike the previous properties, it also declares an additional program \code{pj2} on \cref{line:syn2-2}---\code{pj2} is generated by adding a random conditional to \code{pj1} such that the input/output behavior of the code remains unaffected. For instance, the body of the conditional may just consist of a print statement. On lines~\ref{line:syn2-3}--\ref{line:syn2-4}, the property declares \emph{two} output programs \code{pp1} and \code{pp2}, which are assigned by the \emph{two} model invocations (lines~\ref{line:syn2-6}--\ref{line:syn2-7}). The postcondition checks that the number of loops in \code{pp1} and \code{pp2} matches.

Finally, consider the semantic 2-safety property in \Cref{fig:semantic2}. Here, \code{pj2} is generated from \code{pj1} by renaming a random function parameter (\cref{line:sem2-2}). The precondition expresses that the model should be invoked if the renaming succeeds. The postcondition checks that the return values of the two output programs match provided that both compile---in Python, this means that both programs parse.

In this work, we specified a total of 62 properties---see\apps{sect:appendix-specs}{sect:appendix-specs-search}.

%%----------------------------------------------------------------------------
\subsection{Testing Procedure}
\label{sect:checking-procedure}
%%----------------------------------------------------------------------------

Our testing procedure for these properties is based on the existing \nomos framework~\cite{ChristakisEniser2023}. Internally, the \nomos framework generates a Python test harness for the given model and its specification. The harness tests the model until a user-specified budget is depleted. Specifically, for each budget unit, the harness generates inputs for the model such that any precondition is satisfied, executes the block of Python code in the specification, and checks the postcondition. Finally, the \nomos framework records, processes, and de-duplicates all detected property violations.

The harness essentially implements \emph{metamorphic testing}~\cite{ChenCheung1998,SeguraFraser2016}, which constitutes a natural choice for checking $k$-safety properties. Given an input to a system under test (in our case, a model under test), metamorphic testing transforms the input such that the relation of the corresponding outputs is known. For instance, given a criminal as input to a model that predicts recidivism risk, a metamorphic transformation could increase the criminal's number of priors (keeping all other attributes the same). Then, we know that the recidivism risk of the new criminal should be at least as high as that of the original one. Similarly, a \nomos property for $k > 1$ also describes input transformations, and the expected relation among outputs is the postcondition---see \Cref{fig:examples}.

To support code translation models, we extended \nomos to enable expressing and testing their properties. First, we incorporated two new classes of domain-specific functions, namely, \emph{program-transformation} and \emph{program-inspection} functions. In \Cref{fig:examples}, we use transformation functions \code{addConditional} and \code{renameParam}, and inspection functions \code{numConditionals}, \code{numLoops}, \code{compiles}, and \code{retValues}. In total, we added 7 transformation and 5 inspection functions to express our properties---see\app{sect:appendix-functions}.

LLMs generate their outputs token by token, and by default, the next token is selected greedily by returning the most probable token. The main drawback of this greedy search is that there is a chance of missing high-probability tokens that are hidden behind lower-probability ones, thus generating sub-optimal predictions. A beam size of $N$ tokens alleviates this issue by selecting the most probable $N$ tokens at every step, and in the end, generating $N$ predictions. We, thus, extended \nomos to allow enabling a beam size of $N$, which for a $k$-safety property means that we have $k$ model queries each producing $N$ predictions. When generating the failing tests, \nomos only reports inputs for which all $N^k$ prediction combinations violate the property.

Given that LLMs are expensive to query, we also added a caching mechanism to avoid the cost of asking the same queries repeatedly and thus slowing down our testing procedure. More specifically, due to randomness in the program-transformation functions, we might generate the same inputs for a model under test when checking different (or even the same) properties. We, therefore, cache model queries and the corresponding outputs. Note that this also helps to avoid inconsistent outputs in the case of stochastic models.

Finally, we extended the harness generator to allow users to control different model parameters, like the temperature.

%%----------------------------------------------------------------------------
\subsection{Search Procedure}
\label{sect:search-procedure}
%%----------------------------------------------------------------------------

Being able to check functional properties of code translation models opens up a new use case, namely one where the user only aims to generate a correct translation of a piece of code without necessarily testing the overall model quality. To address this use case, we developed a property-guided search procedure that repeatedly queries a model with slightly different parameters (such as the temperature) to produce alternative and potentially more correct translations with respect to the given properties. In other words, our search procedure takes as input an initial model instance (with user-provided parameters) and searches for model instances (same model but with different parameters) that can satisfy more properties for the particular piece of code.

On a high level, the search procedure repeatedly invokes the testing procedure with mutated model parameters to optimize the number of violated properties. It returns the model output as soon as all properties are satisfied by the current model instance. If all properties cannot be satisfied within a given search budget, it returns the best model output, which results in the fewest violated properties.

\renewcommand\algorithmicthen{}
\renewcommand\algorithmicdo{}

\begin{algorithm}[tb]
\caption{Our search procedure}
\label{alg:algorithm}
\textbf{Input}: properties, program $P$, searchBudget, testBudget,\\
{\color{white}\textbf{Input}:} initModelParams\\
\textbf{Output}: bestTranslation
\begin{algorithmic}[1] %[1] enables line numbers
\STATE params = initModelParams
\STATE \textit{\color{Black!70} // Minimum number of violated properties}
\STATE minVP = +$\infty$
\STATE \textit{\color{Black!70} // Minimum number of total violations}
\STATE minTV = +$\infty$
\STATE bestTranslation = null
\WHILE{searchBudget $>$ 0} \label{line:while}
  \STATE \textit{\color{Black!70} // Current number of violated properties}
  \STATE VP = 0
  \STATE \textit{\color{Black!70} // Current number of total violations}
  \STATE TV = 0
  \STATE \textit{\color{Black!70} /* We run the testing procedure for each property}
  \STATE \textit{\color{Black!70} with the current model parameters */}
  \FORALL{prop \textbf{in} properties} \label{line:forall}
    \STATE \textit{\color{Black!70} /* Function Test returns the number of violations}
    \STATE \textit{\color{Black!70} and the program translation */}
    \STATE v, tr = Test(prop, ${P}$, testBudget, params)
    \STATE TV += v
    \IF{0 $<$ v}
      \STATE VP++
    \ENDIF
  \ENDFOR \label{line:end-forall}
  \IF{VP == 0} \label{line:nobugs}
    \STATE \textbf{return} tr \label{line:end-nobugs}
  \ENDIF
  \IF{(VP $<$ minVP) $\vee$ (VP == minVP $\wedge$ TV $<$ minTV))} \label{line:lex}
    \STATE minVP = VP
    \STATE minTV = TV
    \STATE bestTranslation = tr
  \ENDIF
  \STATE params = Mutate(params) \label{line:params}
  \STATE searchBudget-{}-
\ENDWHILE \label{line:end-while}
\STATE \textbf{return} bestTranslation
\end{algorithmic}
\end{algorithm}

\Cref{alg:algorithm} describes the search procedure more precisely. It takes a set of properties, a program to be translated, a search budget, a test budget, and the initial model parameters; it returns the best program translation found. The while-loop on lines~\ref{line:while}--\ref{line:end-while} iterates until the search budget is depleted. Each iteration runs the testing procedure for all properties with the current model parameters and calculates the number of violated properties and the total number of property violations (lines~\ref{line:forall}--\ref{line:end-forall}). If no properties are violated, the current translation is returned (lines~\ref{line:nobugs}--\ref{line:end-nobugs}). Otherwise, on line~\ref{line:lex}, we use a lexicographic fitness function to minimize the number of violated properties before minimizing the total number of violations. For the next iteration, line~\ref{line:params} mutates the current model parameters. This essentially performs stochastic hill climbing, but more sophisticated optimization techniques could easily be used instead.

For this use case, each $k$-safety property should require a single input ($P_1$) from the user and generate the remaining inputs ($P_2, \ldots, P_k$) automatically. For example, in Figure~\ref{fig:syntactic2}, the user-provided input is \code{pj1}, whereas \code{pj2} is generated from \code{pj1}. Then, our search procedure will optimize the translation of $P_1$.

In addition, note that a violation of a $k$-safety property could be caused by sub-optimal model performance for any of the $k$ model invocations. However, for this use case, the user is only interested in the model behaving as expected for $P_1$, ignoring any violations caused only by its variants $P_2, \ldots, P_k$. This preference can be encoded directly in the property by ensuring that a violation occurs only if the user-provided input is to blame. In particular, each property should only generate equivalent or ``harder" variants of $P_1$, where ``harder" means containing additional code for translation. The postcondition should then express that, if the model succeeds for (potentially harder) $P_2, \ldots, P_k$, then it should also succeed for (potentially easier) $P_1$. When this postcondition is violated, we know that the model output is sub-optimal for the user-provided input.
For instance, to make the property of \Cref{fig:syntactic2} compatible with our search procedure, we could change the postcondition to:
\begin{lstlisting}[style=nomos-inline,numbers=none]
ensures numLoops(pp2, "py") == numLoops(pj2, "java")
  ==> numLoops(pp1, "py") == numLoops(pj1, "java");
\end{lstlisting}

% {\scriptsize
% \begin{verbatim}
% func Search(properties, program, search_budget,
%             checking_budget, init_model_params) : translation {
%   p = init_model_params
%   min_nfp = +inf
%   min_tnf = +inf
%   best_tr = null
%   while search_budget not exceeded {
%     // We run NOMOS for each property
%     // with the current model parameters.
%     // We get the number of failed properties
%     // and the total number of failures.
%     nfp = 0
%     tnf = 0
%     for p in properties {
%       nf, tr = Check(p, {program}, checking_budget, p)
%       tnf += nf
%       if 0 < nf {
%         nfp++
%       }
%     }
%     if nfp == 0 { return tr }
%     if nfp < min_nfp || (nfp == min_nfp && tnf < min_tnf) {
%       min_nfp = nfp
%       min_tnf = tnf
%       best_tr = tr
%     }
%     p = Mutate(p)
%   }
%   return best_tr
% \end{verbatim}
% }

%%----------------------------------------------------------------------------
\section{Evaluation}
\label{sect:evaluation}
%%----------------------------------------------------------------------------

% In this section, we evaluate the effectiveness of our testing and search procedures on four code translation models.

\begin{table*}[t!]
\centering
\scalebox{1.0}{
{\fontsize{9pt}{9pt}\selectfont}
\begin{tabular}{@{}|c|c|c|S[table-format=1.0,table-align-text-pre=false,table-space-text-pre=<]|S[table-format=1.0,table-align-text-pre=false,table-space-text-pre=<]|S[table-format=1.0,table-align-text-pre=false,table-space-text-pre=<]|S[table-format=1.0,table-align-text-pre=false,table-space-text-pre=<]|S[table-format=1.0,table-align-text-pre=false,table-space-text-pre=<]|S[table-format=1.0,table-align-text-pre=false,table-space-text-pre=<]|S[table-format=2.0,table-align-text-pre=false,table-space-text-pre=<,round-mode=places,round-precision=0]|S[table-format=2.0,table-align-text-pre=false,table-space-text-pre=<,round-mode=places,round-precision=0]|S[table-format=2.0,table-align-text-pre=false,table-space-text-pre=<,round-mode=places,round-precision=0]|S[table-format=2.0,table-align-text-pre=false,table-space-text-pre=<,round-mode=places,round-precision=0]|@{}}
\hline
 \multirow{2}{*}{BS} & \multirow{2}{*}{$k$} & \multirow{2}{*}{\diagbox[width=1.3cm]{PT}{PI}} & \multicolumn{2}{c|}{\multirow{2}{*}{\texttt{\textbf{arity}}}} & \multicolumn{2}{c|}{\multirow{2}{*}{\texttt{\textbf{numC/s}}}} & \multicolumn{2}{c|}{\multirow{2}{*}{\texttt{\textbf{numL/s}}}} & \multicolumn{2}{c|}{\multirow{2}{*}{\texttt{\textbf{compiles}}}} & \multicolumn{2}{c|}{\multirow{2}{*}{\texttt{\textbf{retV/s}}}} \\
 & & & \multicolumn{2}{c|}{} & \multicolumn{2}{c|}{} & \multicolumn{2}{c|}{} & \multicolumn{2}{c|}{} & \multicolumn{2}{c|}{} \\ \hline
 \multirow{8}{*}{1} & 1 & -- & 0 & 0 & <1 & <1 & 0 & <1 & 19.2 & 21.8 & 50.8 & 46.8 \\ \cline{2-13}
 & \multirow{6}{*}{2} & \texttt{\textbf{rnmP}} & 0 & <1 & <1 & <1 & 0 & <1 & 14 & 9.2 & 8.3 & 18.9 \\
 &  & \texttt{\textbf{addP}} & 0 & 0 & <1 & <1 & <1 & <1 & 33.7 & 36 & 19.5 & 23.6 \\
 &  & \texttt{\textbf{addC}} & 0 & 0 & <1 & <1 & 0 & <1 & 40.2 & 16.6 & 44.7 & 26 \\
 &  & \texttt{\textbf{chBC}} & 0 & 0 & <1 & <1 & <1 & <1 & 52.2 & 54.2 & \si{-} & \si{-} \\
 &  & \texttt{\textbf{addL}} & 0 & 0 & <1 & <1 & <1 & <1 & 36.8 & 27.6 & 14.8 & 38.9 \\
 &  & \texttt{\textbf{rmL}} & 0 & 0 & 0 & 0 & <1 & 0 & 1.9 & 10.4 & \si{-} & \si{-} \\ \cline{2-13}
 & 3 & \texttt{\textbf{mrg}} & <1 & <1 & <1 & <1 & <1 & <1 & 60.5 & 61.0 & 2.0 & 2.4 \\ \cline{1-13}
 \multirow{8}{*}{3} & 1 & -- & 0 & 0 & 0 & 0 & 0 & <1 & 1 & 1.2 & 4.5 & 4.8 \\ \cline{2-13}
 & \multirow{6}{*}{2} & \texttt{\textbf{rnmP}} & 0 & 0 & <1 & <1 & 0 & 0 & <1 & 4.3 & 0 & <1 \\
 &  & \texttt{\textbf{addP}} & 0 & 0 & 0 & <1 & 0 & <1 & <1 & 1.7 & 8.7 & 6.9 \\
 &  & \texttt{\textbf{addC}} & 0 & 0 & 0 & <1 & 0 & 0 & <1 & 1.6 & 7.5 & 8.2 \\
 &  & \texttt{\textbf{chBC}} & 0 & 0 & <1 & <1 & <1 & <1 & 3.9 & 17 & \si{-} & \si{-} \\
 &  & \texttt{\textbf{addL}} & 0 & 0 & <1 & <1 & <1 & 0 & <1 & 1.1 & <1 & 18.2 \\
 &  & \texttt{\textbf{rmL}} & 0 & 0 & 0 & 0 & 0 & 0 & <1 & <1 & \si{-} & \si{-} \\ \cline{2-13}
 & 3 & \texttt{\textbf{mrg}} & <1 & <1 & <1 & <1 & <1 & <1 & 12.2 & 23.7 & 0 & <1 \\ \hline
\end{tabular}
}
\caption{The percentage of property violations (i.e., unique failing tests / total number of tests x 100\%) detected when running the testing procedure on \tcbase and \tcir for translating from Java to C++. The first column (BS) shows the beam-size parameter, and the second column the value of $k$ of the corresponding $k$-safety property. The third column shows the (abbreviated) program-transformation (PT) functions, which are combined with the (abbreviated) program-inspection (PI) functions of the first row to form the properties in\app{sect:appendix-specs}. We report the percentage of violations under each PI function---the left sub-column shows the percentage of violations for \tcbase and the right for \tcir.}
\label{tab:java-to-cpp-perc}
\end{table*}
% For example, the percentage of violations of the property in \Cref{fig:semantic1} are shown in the $k$=1 rows under \texttt{\textbf{retV/s}}: for BS=1, there is a 51\% of violations for \tcbase and 47\% for \tcir.

\begin{table*}[t!]
\centering
\scalebox{1.0}{
{\fontsize{9pt}{9pt}\selectfont}
\begin{tabular}{@{}|c|c|c|S[table-format=1.0]|S[table-format=1.0,table-align-text-pre=false,table-space-text-pre=<]|S[table-format=1.0,table-align-text-pre=false,table-space-text-pre=<]|S[table-format=1.0,table-align-text-pre=false,table-space-text-pre=<,round-mode=places,round-precision=0]|S[table-format=1.0,table-align-text-pre=false,table-space-text-pre=<,round-mode=places,round-precision=0]|S[table-format=2.0,table-align-text-pre=false,table-space-text-pre=<,round-mode=places,round-precision=0] |S[table-format=1.0,table-align-text-pre=false,table-space-text-pre=<,round-mode=places,round-precision=0]|S[table-format=1.0,table-align-text-pre=false,table-space-text-pre=<,round-mode=places,round-precision=0]|S[table-format=2.0,table-align-text-pre=false,table-space-text-pre=<,round-mode=places,round-precision=0]|S[table-format=2.0,table-align-text-pre=false,table-space-text-pre=<,round-mode=places,round-precision=0]|S[table-format=2.0,table-align-text-pre=false,table-space-text-pre=<,round-mode=places,round-precision=0]|S[table-format=2.0,table-align-text-pre=false,table-space-text-pre=<,round-mode=places,round-precision=0]|S[table-format=2.0,table-align-text-pre=false,table-space-text-pre=<,round-mode=places,round-precision=0]|S[table-format=2.0,table-align-text-pre=false,table-space-text-pre=<,round-mode=places,round-precision=0]|S[table-format=2.0,round-mode=places,round-precision=0]|@{}}
\hline
 \multirow{2}{*}{BS} & \multirow{2}{*}{$k$} & \multirow{2}{*}{\diagbox[width=1.3cm]{PT}{PI}} & \multicolumn{3}{c|}{\multirow{2}{*}{\texttt{\textbf{arity}}}} & \multicolumn{3}{c|}{\multirow{2}{*}{\texttt{\textbf{numC/s}}}} & \multicolumn{3}{c|}{\multirow{2}{*}{\texttt{\textbf{numL/s}}}} & \multicolumn{3}{c|}{\multirow{2}{*}{\texttt{\textbf{compiles}}}} & \multicolumn{3}{c|}{\multirow{2}{*}{\texttt{\textbf{retV/s}}}} \\
 & & & \multicolumn{3}{c|}{} & \multicolumn{3}{c|}{} & \multicolumn{3}{c|}{} & \multicolumn{3}{c|}{} & \multicolumn{3}{c|}{} \\ \hline
  \multirow{8}{*}{1} & 1 & -- & 0 & 0 & <1 & 1.6 & <1 & 2.8 & 1.8 & 3.4 & 2.6 & 20.6 & 16.8 & 6.4 & 42.2 & 47.4 & 58.8 \\ \cline{2-18}
 & \multirow{6}{*}{2} & \texttt{\textbf{rnmP}} & 0 & 0 & 0 & 1.4 & 1.2 & 2.9 & 1.8 & <1 & 38.4 & 7.5 & 7.0 & 4.0 & 5.2 & 5.8 & 17.4 \\
 &  & \texttt{\textbf{addP}} & 0 & 0 & <1 & 2.0 & 1.2 & 3.5 & 3.0 & 1.3 & 3.6 & 7.0 & 8.1 & 46.7 & 5.2 & 4.6 & 14.0 \\
 &  & \texttt{\textbf{addC}} & 0 & <1 & <1 & 1.1 & <1 & 86.4 & 2.3 & 1.4 & 3.1 & 5.2 & 8.6 & 11.9 & 25 & 17.1 & 8.2 \\
 &  & \texttt{\textbf{chBC}} & 0 & 0 & <1 & 2.0 & <1 & 18.6 & 3.2 & 1.1 & 4.0 & 43.3 & 47.3 & 18.6 & \si{-} & \si{-} & \si{-} \\
 &  & \texttt{\textbf{addL}} & 0 & 0 & <1 & <1 & 1.2 & 5.3 & <1 & 1.7 & 42.1 & 8.3 & 8.3 & 11.4 & 24.8 & 26 & 25.1 \\
 &  & \texttt{\textbf{rmL}} & 0 & 0 & <1 & <1 & 0 & 3.2 & 1.3 & <1 & 3.9 & 4.3 & 4.5 & 2.5 & \si{-} & \si{-} & \si{-} \\ \cline{2-18}
 & 3 & \texttt{\textbf{mrg}} & 0 & <1 & <1 & 6.6 & 3.7 & 9.7 & 8.8 & 4.3 & 7 & 39.7 & 45.3 & 19.2 & 5.2 & 6.4 & 20.5 \\ \cline{1-18}
 \multirow{8}{*}{3} & 1 & -- & 0 & 0 & 0 & <1 & 0 & <1 & <1 & <1 & <1 & 3.2 & 2.2 & 1.0 & 5.4 & 6.9 & 11.5 \\ \cline{2-18}
 & \multirow{6}{*}{2} & \texttt{\textbf{rnmP}} & 0 & 0 & 0 & <1 & <1 & <1 & <1 & <1 & <1 & <1 & <1 & <1 & <1 & <1 & 1.3 \\
 &  & \texttt{\textbf{addP}} & 0 & 0 & <1 & 0 & <1 & <1 & <1 & <1 & <1 & <1 & <1 & 3.0 & <1 & <1 & 2.1 \\
 &  & \texttt{\textbf{addC}} & 0 & 0 & 0 & 0 & 0 & 81.5 & <1 & <1 & 1 & <1 & <1 & 7.6 & 15.2 & 8.4 & 3.0 \\
 &  & \texttt{\textbf{chBC}} & 0 & 0 & 0 & <1 & <1 & 11.1 & <1 & <1 & 1.0 & 28.2 & 35.1 & 15.2 & \si{-} & \si{-} & \si{-} \\
 &  & \texttt{\textbf{addL}} & 0 & 0 & 0 & <1 & <1 & <1 & <1 & <1 & 22.6 & <1 & <1 & 3.5 & 13.5 & 16.9 & 18.7 \\
 &  & \texttt{\textbf{rmL}} & 0 & 0 & 0 & 0 & 0 & <1 & <1 & <1 & 3.1 & 2.0 & 1.5 & 1.3 & \si{-} & \si{-} & \si{-} \\ \cline{2-18}
 & 3 & \texttt{\textbf{mrg}} & 0 & 0 & 0 & <1 & <1 & 3.2 & 1.2 & <1 & 2.4 & 14.7 & 21.8 & 55.6 & <1 & 2.4 & 8.8 \\ \hline
\end{tabular}
}
\caption{The percentage of property violations detected when running the testing procedure on \tcbase, \dobf, and \starcoder for translating from Java to Python. Under each PI function, the left sub-column shows the percentage of violations for \tcbase, the middle for \dobf, and the right for \starcoder.}
\label{tab:java-to-py-perc}
\end{table*}
% For example, the percentage of violations of the property in \Cref{fig:semantic2} are shown in the $k$=2, \texttt{\textbf{rnmP}} rows under \texttt{\textbf{retV/s}}: for BS=1, there is a 5\% of violations for \tcbase, 6\% for \dobf, and 17\% for \starcoder.

\begin{table*}[t!]
\centering
{\fontsize{9pt}{9pt}\selectfont}
\begin{tabular}{@{}|c|r|r|r|r|r|r|r|r|r|r|r|r|r|r|r|r|@{}}
\hline
\multirow{3}{*}{BS} & \multicolumn{3}{c|}{\tcbase} & \multicolumn{3}{c|}{\tcir} & \multicolumn{3}{c|}{\tcbase} & \multicolumn{3}{c|}{\dobf} & \multicolumn{3}{c|}{\starcoder} \\
& \multicolumn{3}{c|}{Java-C++} & \multicolumn{3}{c|}{Java-C++} & \multicolumn{3}{c|}{Java-Python} & \multicolumn{3}{c|}{Java-Python} & \multicolumn{3}{c|}{Java-Python} \\
& TV & VP & VSP & TV & VP & VSP & TV & VP & VSP & TV & VP & VSP & TV & VP & VSP \\ \hline
1 & 8663 & 27 & 13 & 8603 & 30 & 16 & 5777 & 30 & 16 & 5564 & 31 & 17 & 11213 & 37 & 23 \\ \hline
3 & 1035 & 20 & 8 & 2326 & 25 & 11 & 2240 & 27 & 13 & 2535 & 26 & 13 & 6611 & 31 & 17 \\ \hline
\end{tabular}
\caption{The total number of violations (TV), the number of violated properties (VP), and the number of violated syntactic properties (VSP) detected when running the testing procedure on all models.}
\label{tab:props}
\end{table*}

%%----------------------------------------------------------------------------
\subsection{Experimental Setup}
\label{sect:exp_setup}
%%----------------------------------------------------------------------------

\textbf{Models.} For our experiments, we use the pre-trained models \tcbase~\cite{RoziereLachaux2020}, \dobf~\cite{LachauxRoziere2021}, \tcir~\cite{SzafraniecRoziere2023}, and \starcoder~\cite{LiAllal2023}. The first three expect a function as input and predict the corresponding function in the output language. For \starcoder, we use completion mode and send queries that provide an input function and request the output function to be completed.
We evaluate \tcbase for both Java-to-C++ and Java-to-Python translations. \tcir is evaluated for Java to C++ (it was not trained on Python) and \dobf for Java to Python (it was not trained on C++). We evaluate \starcoder for Java to Python since it was fine-tuned for Python.

\textbf{Program benchmarks.} We use the benchmark set introduced in \tcbase~\cite{RoziereLachaux2020}, including 545 solutions to LeetCode problems implemented in Java, C++, and Python.
On average, each program comes with ca. 10 tests that specify the expected output values for given input values. (These are the input values used by \code{retValues}.)

\textbf{Parameters.} We run the testing procedure with beam size 1 and 3 and set the temperature to 0.1. For the search, the beam size is 1, and we mutate the model temperature, which is initially 0.1.

%%----------------------------------------------------------------------------
\subsection{Results for Testing Procedure}
\label{sect:results_testing}
%%----------------------------------------------------------------------------

We evaluate our testing procedure by checking a total of 38 properties found in\app{sect:appendix-specs}.

We introduced many inspection and transformation functions; we now provide a summary of the remaining ones (see\app{sect:appendix-functions} for details). The \code{arity} inspection function returns the number of parameters of a given function. The \code{addParam} transformation function adds a random parameter to the function without affecting its input/output behavior, and \code{addLoop} adds a random for-loop. The \code{rmLoop} function removes a random for-loop from the function, while \code{chBranchCond} randomly changes the branch condition of an if- or switch-statement. Both \code{rmLoop} and \code{chBranchCond} may change the input/output behavior of the original code and are thus not used in semantic properties. Finally, \code{merge} merges two functions by executing one of them depending on an additional Boolean argument.

In our experiments, we set the testing budget to 500 for all 1-safety properties and to 2500 for all other properties. All 545 programs from our benchmark set may be used as inputs when testing the models. Note that the testing budget for 1-safety properties is lower since these properties only select a single program from the corpus and invoke the model under test on this program; thus, a higher budget would unlikely result in a significantly higher number of unique violations.

Next, we present our results for the testing procedure organized in the following five research questions (RQs).

\textbf{RQ1. Is the testing procedure effective in detecting property violations?} Tables~\ref{tab:java-to-cpp-perc} and \ref{tab:java-to-py-perc} show the percentages of (unique) property violations for all models. The absolute numbers are included in\app{sect:appendix-results} but are summarized in \Cref{tab:props}.

When using a beam size of 1, our testing procedure finds between 27 (\tcbase for Java to C++) and 37 (\starcoder for Java to Python) violated properties (out of 38). The number of total property violations ranges from 5564 (\dobf for Java to Python) to 11213 (\starcoder for Java to Python). Our testing procedure is, therefore, highly effective in detecting violations.

When increasing the beam size to 3, we observe a reduced number of violations and violated properties. In particular, the number of violated properties decreases by between 3 (\tcbase for Java to Python) and 7 (\tcbase for Java to C++). This confirms that increasing the beam size can improve the quality of the translations.

\textbf{RQ2. How do models perform with respect to different properties?}
As shown in the tables, the models generally perform better for purely syntactic properties, i.e., using the \code{arity}, \code{numConditionals}, and \code{numLoops} inspection functions.
In particular, for a beam size of 1, the number of violated syntactic properties ranges from 13 (\tcbase for Java to C++) to 23 (\starcoder for Java to C++) (out of 24), whereas the number of other violated properties (i.e., using \code{compiles} and \code{retValues}) is 14 (out of 14) for all models.
For a beam size of 3, the number of violated syntactic properties decreases for all models, and more specifically, by between 3 (\tcbase for Java to Python) and 6 (\starcoder for Java to Python). In contrast, the number of other violated properties only decreases by 2 for \tcbase for Java to C++ and by 1 for \dobf for Java to Python.

\textbf{RQ3. How does \tcbase perform with respect to the target language?}
We only evaluate \tcbase for two target languages, namely C++ and Python. It generally performs better for C++, where it violates 27 (out of 38) properties for a beam size of 1 and 20 for a beam size of 3. For Python, it violates 30 and 27 properties, respectively. This not unexpected since C++ is more similar to Java than Python. Interestingly however, for \tcbase for C++, we detect a total of 6084 violations of \code{compiles} properties in contrast to 2996 violations for Python (for a beam size of 1). Again, this is not unexpected since these properties only check program parsing for Python.

\textbf{RQ4. How do translation models perform in comparison to the general-purpose model \starcoder?}
We found significantly more property violations for \starcoder than for the specialized translation models. In particular, for a beam size of 1, we detected 37 (out of 38) violated properties and 11213 total violations for \starcoder, 30 violated properties and 5777 total violations for \tcbase, and 31 violated properties and 5564 total violations for \dobf. This is not surprising given that the specialized models are specifically trained for translation.

\textbf{RQ5. What is the average running time for checking a property on a given model?}
The average running time (including model inference and test harness execution) for checking a property ranges between 1.3s (\dobf for Java to Python) and 42.9s (\starcoder for Java to Python) for beam size 1. When increasing the beam size to 3, the running time increases slightly for all models (for instance, to 3.3s for \dobf for Java to Python and to 51.3s for \starcoder for Java to Python). We include the average running time for all models in\app{sect:appendix-results}.

These experiments were run on cluster machines with A100 Nvidia Tesla GPUs and Intel Xeon Gold 5317 CPUs, running Debian GNU/Linux 11. Each GPU has 80GB memory allowing to host the larger \starcoder model. 

%%----------------------------------------------------------------------------
\subsection{Results for Search Procedure}
\label{sect:results_search}
%%----------------------------------------------------------------------------

\begin{figure}[t!]
\centering
\begin{subfigure}[b]{\linewidth}
    \centering
    \includegraphics[clip, width=0.90\linewidth]{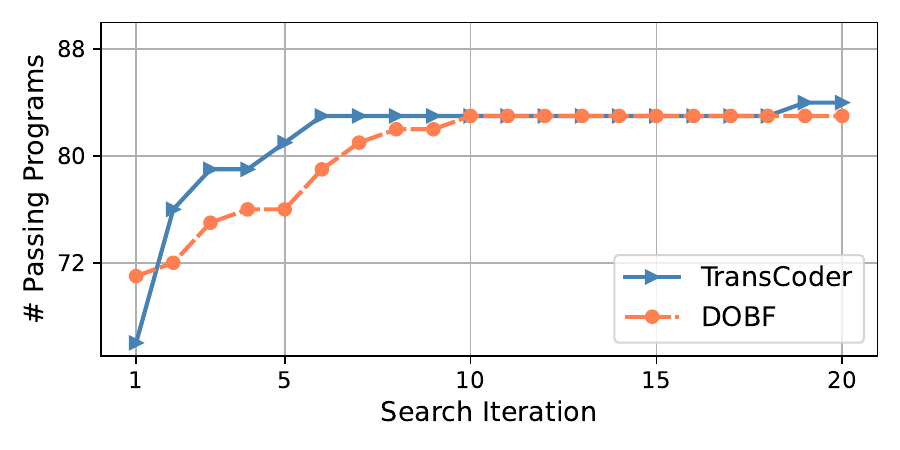}
    \label{fig:tc-dobf-incr}
\end{subfigure}
\begin{subfigure}[t!]{\linewidth}
    \centering
    \includegraphics[clip, width=0.90\linewidth]{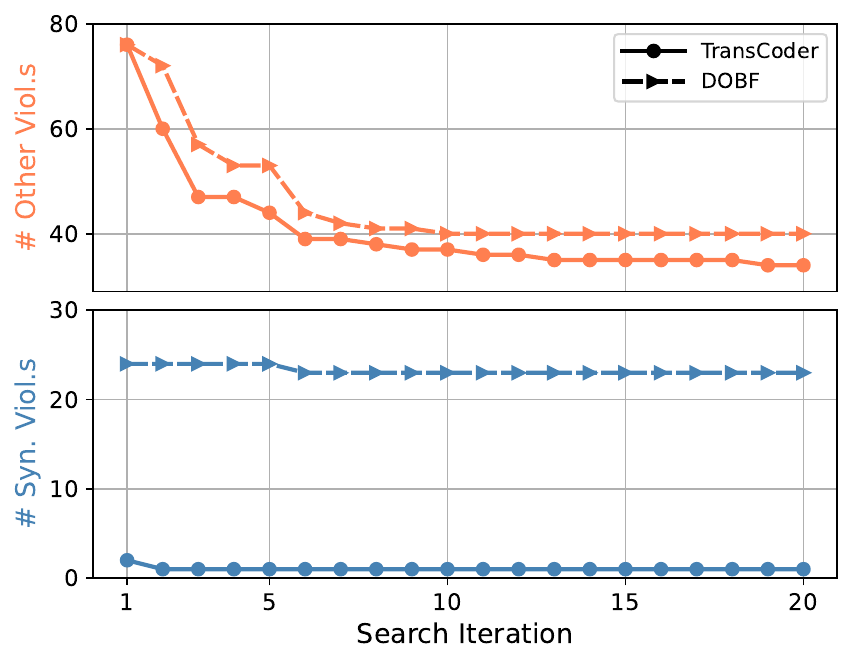}
    \label{fig:tc-dobf-sem-syn}
\end{subfigure}
\caption{The number of passing programs (top) and the number of syntactic and other property violations (bottom) as the number of search iterations increases from 1 to 20.}
\label{fig:tc-dobf}
\end{figure}

We evaluate the effectiveness of our search procedure for \tcbase and \dobf by generating Java-to-Python translations for 100 randomly selected programs from our benchmark set. We check 24 search properties (see\app{sect:appendix-specs-search}) for each translation.

We set the search budget to 20 and the testing budget to 50 (see \Cref{alg:algorithm}). We use relatively small budgets to keep the running time small. We start the search with an initial temperature of 0.1 and mutate it at every iteration by adding Gaussian noise with $\mu = 0$ and $\sigma^2 = 0.01$.

% Next, we present our results for the search procedure organized in the following three research questions.

\textbf{RQ1. Is the search procedure effective in finding better translations?}
\Cref{fig:tc-dobf} (top) shows the number of ``passing programs'' (i.e., programs for which the testing procedure reports no violations) out of 100 along the y-axis as we increase the number of search iterations from 1 to 20 along the x-axis. As shown in the figure, the search significantly increases the number of passing programs (from 66 to 84 for \tcbase and from 71 to 83 for \dobf). It also reduces the number of violated properties for the final translation by almost half (from 78 to 35 for \tcbase and from 100 to 63 for \dobf). Our search procedure is, therefore, effective in improving the quality of the translation with respect to the number of violated properties.

\textbf{RQ2. How long does the search take?}
As shown in \Cref{fig:tc-dobf} (top), exploring only a few (e.g., 6) model instances leads to significant improvements. The mean number of search iterations is 4.5 for \tcbase and 4.7 for \dobf.

\textbf{RQ3. How does the search affect the number of violations of syntactic versus other properties?}
\Cref{fig:tc-dobf} (bottom) shows how the search affects the number of violations of syntactic and other properties (i.e., using \code{compiles} and \code{retValues}). Interestingly, the search helps the most in decreasing the number of violations of other properties, which are the hardest to satisfy.

%%----------------------------------------------------------------------------
\section{Conclusion}
\label{sect:conclusion}
%%----------------------------------------------------------------------------

In this paper, we introduced the first approach for automatically testing user-provided, functional properties of code translation models. 
We extended the \nomos framework with domain-specific functions to formalize a broad set of 38 properties, which we evaluated by testing four popular models. 
We also introduced a property-guided search procedure that aims to optimize the model output based on the number of violated properties.

In future work, we plan to transfer this idea to other settings to effectively enforce quality criteria on model outputs.

%%----------------------------------------------------------------------------
\iftoggle{longversion}{
\appendix
\onecolumn
%%----------------------------------------------------------------------------
\section{New \nomos Domain-Specific Functions}
\label{sect:appendix-functions}
%%----------------------------------------------------------------------------

%%----------------------------------------------------------------------------
\subsection{Program-Transformation Functions}
%%----------------------------------------------------------------------------

\noindent
\textbf{\code{addConditional}:} Adds a random conditional in a given function without affecting its input/output behavior.

\noindent
\textbf{\code{addLoop}:} Adds a random for-loop in a given function without affecting its input/output behavior.

\noindent
\textbf{\code{addParam}:} Adds a random parameter in a given function without affecting its input/output behavior.

\noindent
\textbf{\code{chBranchCond}:} Replaces the condition of a random if- or switch-statement in a given function. The new condition is generated randomly, and therefore, the input/output behavior of the transformed function may not match the input/output behavior of the original function.

\noindent
\textbf{\code{merge}:} Merges the bodies of two given functions into a single function. The new function takes as input a Boolean parameter in addition to the parameters of the two given functions. The body of the new function contains a conditional that branches on the Boolean parameter, and each branch executes one of the given functions.

\noindent
\textbf{\code{renameParam}:} Renames a random parameter in a given function without affecting its input/output behavior.

\noindent
\textbf{\code{rmLoop}:} Removes a random for-loop from a given function. As a result, the input/output behavior of the transformed function may not match the input/output behavior of the original function.

%%----------------------------------------------------------------------------
\subsection{Program-Inspection Functions}
%%----------------------------------------------------------------------------

\noindent
\textbf{\code{arity}:} Returns the number of parameters of a given function.

\noindent
\textbf{\code{compiles}:} Returns true if the given function compiles, otherwise it returns false.

\noindent
\textbf{\code{numConditionals}:} Returns the number of conditionals in a given function.

\noindent
\textbf{\code{numLoops}:} Returns the number of loops in a given function.

\noindent
\textbf{\code{retValues}:} Returns the output values of a given function. The inputs could be randomly generated or imported from an existing test suite of the model under test.

%%----------------------------------------------------------------------------
\section{Specifications for Testing}
\label{sect:appendix-specs}
%%----------------------------------------------------------------------------

We name each of our properties based on the program-transformation and program-inspection functions it uses as follows: $<$program-transformation function$>|<$program-inspection function$>$. Note that a program-transformation function is optional. We describe how each function is implemented in \Cref{sect:appendix-functions}.

%%----------------------------------------------------------------------------
\subsection{1-Safety Properties}
%%----------------------------------------------------------------------------

\noindent
\textbf{arity.} The arity of the output function should be equal to the arity of the input function.

\noindent
\textbf{numConditionals.} The number of conditionals in the output function should be equal to the number of conditionals in the input function.

\noindent
\textbf{numLoops.} The number of loops in the output function should be equal to the number of loops in the input function.

\noindent
\textbf{compiles.} If the input function compiles (resp. does not compile), then the output function should also (resp. not) compile. In other words, compilation should be preserved among the input and output functions.

\noindent
\textbf{retValues.} Given an input function that compiles, if the output function also compiles, then the return values of the output function should be equal to the return values of the input function.

%%----------------------------------------------------------------------------
\subsection{2-Safety Properties}
%%----------------------------------------------------------------------------

%% renameParam
\noindent
\textbf{renameParam$|$arity.} Given input function $i_1$, input function $i_2$ is created by renaming a random parameter in $i_1$. The arity of their corresponding output functions should be equal.

\noindent
\textbf{renameParam$|$numConditionals.} Given input function $i_1$, input function $i_2$ is created by renaming a random parameter in $i_1$. The number of conditionals in their corresponding output functions should be equal.

\noindent
\textbf{renameParam$|$numLoops.} Given input function $i_1$, input function $i_2$ is created by renaming a random parameter in $i_1$. The number of loops in their corresponding output functions should be equal.

\noindent
\textbf{renameParam$|$compiles.} Given input function $i_1$, input function $i_2$ is created by renaming a random parameter in $i_1$. Compilation should be preserved among their corresponding output functions.

\noindent
\textbf{renameParam$|$retValues.} Given input function $i_1$, input function $i_2$ is created by renaming a random parameter in $i_1$. If their corresponding output functions compile, then their return values should be equal. \newline

%% addParam
\noindent
\textbf{addParam$|$arity.} Given input function $i_1$, input function $i_2$ is created by adding a random parameter in $i_1$. Assuming that $o_1$ and $o_2$ are their corresponding output functions, the arity of $o_2$ should be greater than the arity of $o_1$.

\noindent
\textbf{addParam$|$numConditionals.} Given input function $i_1$, input function $i_2$ is created by adding a random parameter in $i_1$. The number of conditionals in their corresponding output functions should be equal.

\noindent
\textbf{addParam$|$numLoops.} Given input function $i_1$, input function $i_2$ is created by adding a random parameter in $i_1$. The number of loops in their corresponding output functions should be equal.

\noindent
\textbf{addParam$|$compiles.} Given input function $i_1$, input function $i_2$ is created by adding a random parameter in $i_1$. Compilation should be preserved among their corresponding output functions.

\noindent
\textbf{addParam$|$retValues.} Given input function $i_1$, input function $i_2$ is created by adding a random parameter in $i_1$. If their corresponding output functions compile, then their return values should be equal. \newline

%% addConditional
\noindent
\textbf{addConditional$|$arity.} Given input function $i_1$, input function $i_2$ is created by adding a random conditional in $i_1$. The arity of their corresponding output functions should be equal.

\noindent
\textbf{addConditional$|$numConditionals.} Given input function $i_1$, input function $i_2$ is created by adding a random conditional in $i_1$. Assuming that $o_1$ and $o_2$ are their corresponding output functions, the number of conditionals in $o_2$ should be greater than the number of conditionals in $o_1$.

\noindent
\textbf{addConditional$|$numLoops.} Given input function $i_1$, input function $i_2$ is created by adding a random conditional in $i_1$. The number of loops in their corresponding output functions should be equal.

\noindent
\textbf{addConditional$|$compiles.} Given input function $i_1$, input function $i_2$ is created by adding a random conditional in $i_1$. Compilation should be preserved among their corresponding output functions.

\noindent
\textbf{addConditional$|$retValues.} Given input function $i_1$, input function $i_2$ is created by adding a random conditional in $i_1$. If their corresponding output functions compile, then their return values should be equal. \newline

%% chBranchCond
\noindent
\textbf{chBranchCond$|$arity.} Given input function $i_1$, input function $i_2$ is created by replacing the condition of a random if- or switch-statement in $i_1$ with a random condition. The arity of their corresponding output functions should be equal.

\noindent
\textbf{chBranchCond$|$numConditionals.} Given input function $i_1$, input function $i_2$ is created by replacing the condition of a random if- or switch-statement in $i_1$ with a random condition. The number of conditionals in their corresponding output functions should be equal.

\noindent
\textbf{chBranchCond$|$numLoops.} Given input function $i_1$, input function $i_2$ is created by replacing the condition of a random if- or switch-statement in $i_1$ with a random condition. The number of loops in their corresponding output functions should be equal.

\noindent
\textbf{chBranchCond$|$compiles.} Given input function $i_1$, input function $i_2$ is created by replacing the condition of a random if- or switch-statement in $i_1$ with a random condition. Compilation should be preserved among their corresponding output functions. \newline

%% addLoop
\noindent
\textbf{addLoop$|$arity.} Given input function $i_1$, input function $i_2$ is created by adding a random for-loop in $i_1$. The arity of their corresponding output functions should be equal.

\noindent
\textbf{addLoop$|$numConditionals.} Given input function $i_1$, input function $i_2$ is created by adding a random for-loop in $i_1$. The number of conditionals in their corresponding output functions should be equal.

\noindent
\textbf{addLoop$|$numLoops.} Given input function $i_1$, input function $i_2$ is created by adding a random for-loop in $i_1$. Assuming that $o_1$ and $o_2$ are their corresponding output functions, the number of loops in $o_2$ should be greater than the number of loops in $o_1$.

\noindent
\textbf{addLoop$|$compiles.} Given input function $i_1$, input function $i_2$ is created by adding a random for-loop in $i_1$. Compilation should be preserved among their corresponding output functions.

\noindent
\textbf{addLoop$|$retValues.} Given input function $i_1$, input function $i_2$ is created by adding a random for-loop in $i_1$. If their corresponding output functions compile, then their return values should be equal. \newline

%% rmLoop
\noindent
\textbf{rmLoop$|$arity.} Given input function $i_1$, input function $i_2$ is created by removing a random for-loop in $i_1$. The arity of their corresponding output functions should be equal.

\noindent
\textbf{rmLoop$|$numConditionals.} Given input function $i_1$, input function $i_2$ is created by removing a random for-loop in $i_1$. The number of conditionals in their corresponding output functions should be equal.

\noindent
\textbf{rmLoop$|$numLoops.} Given input function $i_1$, input function $i_2$ is created by removing a random for-loop in $i_1$. Assuming that $o_1$ and $o_2$ are their corresponding output functions, the number of loops in $o_2$ should be less than the number of loops in $o_1$.

\noindent
\textbf{rmLoop$|$compiles.} Given input function $i_1$, input function $i_2$ is created by removing a random for-loop in $i_1$. Compilation should be preserved among their corresponding output functions.

%%----------------------------------------------------------------------------
\subsection{3-Safety Properties}
%%----------------------------------------------------------------------------

\noindent
\textbf{merge$|$arity.} Given input functions $i_1$ and $i_2$, input function $i_3$ is created by merging $i_1$ and $i_2$. Assuming that $o_1$, $o_2$, and $o_3$ are their corresponding output functions, the arity of $o_3$ should be equal to the arity of $o_1$ plus the arity of $o_2$ plus $1$.

\noindent
\textbf{merge$|$numConditionals.} Given input functions $i_1$ and $i_2$, input function $i_3$ is created by merging $i_1$ and $i_2$. Assuming that $o_1$, $o_2$, and $o_3$ are their corresponding output functions, the number of conditionals in $o_3$ should be equal to the number of conditionals in $o_1$ plus the number of conditionals in $o_2$ plus $1$.

\noindent
\textbf{merge$|$numLoops.} Given input functions $i_1$ and $i_2$, input function $i_3$ is created by merging $i_1$ and $i_2$. Assuming that $o_1$, $o_2$, and $o_3$ are their corresponding output functions, the number of loops in $o_3$ should be equal to the number of loops in $o_1$ plus the number of loops in $o_2$.

\noindent
\textbf{merge$|$compiles.} Given input functions $i_1$ and $i_2$, input function $i_3$ is created by merging $i_1$ and $i_2$. Assuming that $o_1$, $o_2$, and $o_3$ are their corresponding output functions, if $o_1$ and $o_2$ compile, then $o_3$ should also compile.

\noindent
\textbf{merge$|$retValues.} Given input functions $i_1$ and $i_2$, input function $i_3$ is created by merging $i_1$ and $i_2$. Assuming that $o_1$, $o_2$, and $o_3$ are their corresponding output functions, if $o_1$, $o_2$, and $o_3$ compile, then the return values of $o_3$ should be equal either to the return values of $o_1$ or $o_2$.

%%----------------------------------------------------------------------------
\section{Specifications for Search}
\label{sect:appendix-specs-search}
%%----------------------------------------------------------------------------

For our search procedure, we only specify 2-safety properties. We again name our properties based on the program-transformation and program-inspection functions they use and add an ``S" at the end to differentiate them from the properties written for testing: $<$program-transformation function$>|<$program-inspection function$>|$ S. \newline

%% renameParam
\noindent
\textbf{renameParam$|$arity$|$S.} Given input function $i_1$, input function $i_2$ is created by renaming a random parameter in $i_1$. Assuming that $o_1$ and $o_2$ are their corresponding output functions, if the arity of $o_2$ is equal to the arity of $i_2$, then the arity of $o_1$ should also be equal to the arity of $i_1$.

\noindent
\textbf{renameParam$|$numConditionals$|$S.} Given input function $i_1$, input function $i_2$ is created by renaming a random parameter in $i_1$. Assuming that $o_1$ and $o_2$ are their corresponding output functions, if the number of conditionals in $o_2$ is equal to the number of conditionals in $i_2$, then the number of conditionals in $o_1$ should also be equal to the number of conditionals in $i_1$.

\noindent
\textbf{renameParam$|$numLoops$|$S.} Given input function $i_1$, input function $i_2$ is created by renaming a random parameter in $i_1$. Assuming that $o_1$ and $o_2$ are their corresponding output functions, if the number of loops in $o_2$ is equal to the number of loops in $i_2$, then the number of loops in $o_1$ should also be equal to the number of loops in $i_1$.

\noindent
\textbf{renameParam$|$compiles$|$S.} Given input function $i_1$, input function $i_2$ is created by renaming a random parameter in $i_1$. Assuming that $o_1$ and $o_2$ are their corresponding output functions, if $o_2$ compiles, then $o_1$ should also compile.

\noindent
\textbf{renameParam$|$retValues$|$S.} Given input function $i_1$, input function $i_2$ is created by renaming a random parameter in $i_1$. Assuming that $o_1$ and $o_2$ are their corresponding, \emph{compiling} output functions, if the return values of $o_2$ are equal to the return values of $i_2$, then the return values of $o_1$ should also be equal to the return values of $i_1$. \newline

%% addParam
\noindent
\textbf{addParam$|$arity$|$S.} Given input function $i_1$, input function $i_2$ is created by adding a random parameter in $i_1$. Assuming that $o_1$ and $o_2$ are their corresponding output functions, if the arity of $o_2$ is equal to the arity of $i_2$, then the arity of $o_1$ should also be equal to the arity of $i_1$.

\noindent
\textbf{addParam$|$numConditionals$|$S.} Given input function $i_1$, input function $i_2$ is created by adding a random parameter in $i_1$. Assuming that $o_1$ and $o_2$ are their corresponding output functions, if the number of conditionals in $o_2$ is equal to the number of conditionals in $i_2$, then the number of conditionals in $o_1$ should also be equal to the number of conditionals in $i_1$.

\noindent
\textbf{addParam$|$numLoops$|$S.} Given input function $i_1$, input function $i_2$ is created by adding a random parameter in $i_1$. Assuming that $o_1$ and $o_2$ are their corresponding output functions, if the number of loops in $o_2$ is equal to the number of loops in $i_2$, then the number of loops in $o_1$ should also be equal to the number of loops in $i_1$.

\noindent
\textbf{addParam$|$compiles$|$S.} Given input function $i_1$, input function $i_2$ is created by adding a random parameter in $i_1$. Assuming that $o_1$ and $o_2$ are their corresponding output functions, if $o_2$ compiles, then $o_1$ should also compile.

\noindent
\textbf{addParam$|$retValues$|$S.} Given input function $i_1$, input function $i_2$ is created by adding a random parameter in $i_1$. Assuming that $o_1$ and $o_2$ are their corresponding, \emph{compiling} output functions, if the return values of $o_2$ are equal to the return values of $i_2$, then the return values of $o_1$ should also be equal to the return values of $i_1$. \newline

%% addConditional
\noindent
\textbf{addConditional$|$arity$|$S.} Given input function $i_1$, input function $i_2$ is created by adding a random conditional in $i_1$. Assuming that $o_1$ and $o_2$ are their corresponding output functions, if the arity of $o_2$ is equal to the arity of $i_2$, then the arity of $o_1$ should also be equal to the arity of $i_1$.

\noindent
\textbf{addConditional$|$numConditionals$|$S.} Given input function $i_1$, input function $i_2$ is created by adding a random conditional in $i_1$. Assuming that $o_1$ and $o_2$ are their corresponding output functions, if the number of conditionals in $o_2$ is equal to the number of conditionals in $i_2$, then the number of conditionals in $o_1$ should also be equal to the number of conditionals in $i_1$.

\noindent
\textbf{addConditional$|$numLoops$|$S.} Given input function $i_1$, input function $i_2$ is created by adding a random conditional in $i_1$. Assuming that $o_1$ and $o_2$ are their corresponding output functions, if the number of loops in $o_2$ is equal to the number of loops in $i_2$, then the number of loops in $o_1$ should also be equal to the number of loops in $i_1$.

\noindent
\textbf{addConditional$|$compiles$|$S.} Given input function $i_1$, input function $i_2$ is created by adding a random conditional in $i_1$. Assuming that $o_1$ and $o_2$ are their corresponding output functions, if $o_2$ compiles, then $o_1$ should also compile.

\noindent
\textbf{addConditional$|$retValues$|$S.} Given input function $i_1$, input function $i_2$ is created by adding a random conditional in $i_1$. Assuming that $o_1$ and $o_2$ are their corresponding, \emph{compiling} output functions, if the return values of $o_2$ are equal to the return values of $i_2$, then the return values of $o_1$ should also be equal to the return values of $i_1$. \newline

%% chBranchCond
\noindent
\textbf{chBranchCond$|$arity$|$S.} Given input function $i_1$, input function $i_2$ is created by replacing the condition of a random if- or switch-statement in $i_1$ with a random condition. Assuming that $o_1$ and $o_2$ are their corresponding output functions, if the arity of $o_2$ is equal to the arity of $i_2$, then the arity of $o_1$ should also be equal to the arity of $i_1$.

\noindent
\textbf{chBranchCond$|$numConditionals$|$S.} Given input function $i_1$, input function $i_2$ is created by replacing the condition of a random if- or switch-statement in $i_1$ with a random condition. Assuming that $o_1$ and $o_2$ are their corresponding output functions, if the number of conditionals in $o_2$ is equal to the number of conditionals in $i_2$, then the number of conditionals in $o_1$ should also be equal to the number of conditionals in $i_1$.

\noindent
\textbf{chBranchCond$|$numLoops$|$S.} Given input function $i_1$, input function $i_2$ is created by replacing the condition of a random if- or switch-statement in $i_1$ with a random condition. Assuming that $o_1$ and $o_2$ are their corresponding output functions, if the number of loops in $o_2$ is equal to the number of loops in $i_2$, then the number of loops in $o_1$ should also be equal to the number of loops in $i_1$.

\noindent
\textbf{chBranchCond$|$compiles$|$S.} Given input function $i_1$, input function $i_2$ is created by replacing the condition of a random if- or switch-statement in $i_1$ with a random condition. Assuming that $o_1$ and $o_2$ are their corresponding output functions, if $o_2$ compiles, then $o_1$ should also compile. \newline

%% addLoop
\noindent
\textbf{addLoop$|$arity$|$S.} Given input function $i_1$, input function $i_2$ is created by adding a random for-loop in $i_1$. Assuming that $o_1$ and $o_2$ are their corresponding output functions, if the arity of $o_2$ is equal to the arity of $i_2$, then the arity of $o_1$ should also be equal to the arity of $i_1$.

\noindent
\textbf{addLoop$|$numConditionals$|$S.} Given input function $i_1$, input function $i_2$ is created by adding a random for-loop in $i_1$. Assuming that $o_1$ and $o_2$ are their corresponding output functions, if the number of conditionals in $o_2$ is equal to the number of conditionals in $i_2$, then the number of conditionals in $o_1$ should also be equal to the number of conditionals in $i_1$.

\noindent
\textbf{addLoop$|$numLoops$|$S.} Given input function $i_1$, input function $i_2$ is created by adding a random for-loop in $i_1$. Assuming that $o_1$ and $o_2$ are their corresponding output functions, if the number of loops in $o_2$ is equal to the number of loops in $i_2$, then the number of loops in $o_1$ should also be equal to the number of loops in $i_1$.

\noindent
\textbf{addLoop$|$compiles$|$S.} Given input function $i_1$, input function $i_2$ is created by adding a random for-loop in $i_1$. Assuming that $o_1$ and $o_2$ are their corresponding output functions, if $o_2$ compiles, then $o_1$ should also compile.

\noindent
\textbf{addLoop$|$retValues$|$S.} Given input function $i_1$, input function $i_2$ is created by adding a random for-loop in $i_1$. Assuming that $o_1$ and $o_2$ are their corresponding, \emph{compiling} output functions, if the return values of $o_2$ are equal to the return values of $i_2$, then the return values of $o_1$ should also be equal to the return values of $i_1$.

%%----------------------------------------------------------------------------
\section{Experimental Results}
\label{sect:appendix-results}
%%----------------------------------------------------------------------------

Tables~\ref{tab:java-to-cpp}, \ref{tab:java-to-py}, and \ref{tab:time} present our additional experimental results.

\begin{table*}[ht!]
\centering
{\fontsize{9pt}{9pt}\selectfont}
\begin{tabular}{@{}|c|c|c|S[table-format=1.0]|S[table-format=1.0]|S[table-format=2.0]|S[table-format=2.0]|S[table-format=2.0]|S[table-format=2.0]|S[table-format=4.0]|S[table-format=4.0]|S[table-format=4.0]|S[table-format=3.0]|@{}}
\hline
 \multirow{2}{*}{BS} & \multirow{2}{*}{$k$} & \multirow{2}{*}{\diagbox[width=1.3cm]{PT}{PI}} & \multicolumn{2}{c|}{\multirow{2}{*}{\texttt{\textbf{arity}}}} & \multicolumn{2}{c|}{\multirow{2}{*}{\texttt{\textbf{numC/s}}}} & \multicolumn{2}{c|}{\multirow{2}{*}{\texttt{\textbf{numL/s}}}} & \multicolumn{2}{c|}{\multirow{2}{*}{\texttt{\textbf{compiles}}}} & \multicolumn{2}{c|}{\multirow{2}{*}{\texttt{\textbf{retV/s}}}} \\
 & & & \multicolumn{2}{c|}{} & \multicolumn{2}{c|}{} & \multicolumn{2}{c|}{} & \multicolumn{2}{c|}{} & \multicolumn{2}{c|}{} \\ \hline
 \multirow{8}{*}{1} & 1 & -- & 0 & 0 & 1 & 1 & 0 & 2 & 96 & 109 & 254 & 234 \\ \cline{2-13}
 & \multirow{6}{*}{2} & \texttt{\textbf{rnmP}} & 0 & 1 & 3 & 3 & 0 & 7 & 351 & 232 & 209 & 474 \\
 &  & \texttt{\textbf{addP}} & 0 & 0 & 4 & 11 & 3 & 19 & 843 & 902 & 489 & 591 \\
 &  & \texttt{\textbf{addC}} & 0 & 0 & 4 & 1 & 0 & 11 & 1006 & 416 & 1119 & 652 \\
 &  & \texttt{\textbf{chBC}} & 0 & 0 & 15 & 8 & 9 & 10 & 1305 & 1355 & \si{-} & \si{-} \\
 &  & \texttt{\textbf{addL}} & 0 & 0 & 8 & 9 & 2 & 3 & 920 & 692 & 371 & 974 \\
 &  & \texttt{\textbf{rmL}} & 0 & 0 & 0 & 0 & 3 & 0 & 49 & 261 & \si{-} & \si{-} \\ \cline{2-13}
 & 3 & \texttt{\textbf{mrg}} & 5 & 8 & 17 & 9 & 11 & 23 & 1514 & 1525 & 52 & 60\\ \cline{1-13}
 \multirow{8}{*}{3} & 1 & -- & 0 & 0 & 0 & 0 & 0 & 1 & 25 & 30 & 113 & 122 \\ \cline{2-13}
 & \multirow{6}{*}{2} & \texttt{\textbf{rnmP}} & 0 & 0 & 1 & 1 & 0 & 0 & 1 & 109 & 0 & 8 \\
 &  & \texttt{\textbf{addP}} & 0 & 0 & 0 & 6 & 0 & 1 & 7 & 44 & 219 & 174 \\
 &  & \texttt{\textbf{addC}} & 0 & 0 & 0 & 4 & 0 & 0 & 14 & 41 & 189 & 206 \\
 &  & \texttt{\textbf{chBC}} & 0 & 0 & 13 & 13 & 9 & 10 & 98 & 427 & \si{-} & \si{-} \\
 &  & \texttt{\textbf{addL}} & 0 & 0 & 5 & 5 & 1 & 0 & 7 & 29 & 2 & 457 \\
 &  & \texttt{\textbf{rmL}} & 0 & 0 & 0 & 0 & 0 & 0 & 6 & 13 & \si{-} & \si{-} \\ \cline{2-13}
 & 3 & \texttt{\textbf{mrg}} & 4 & 3 & 10 & 10 & 6 & 7 & 305 & 594 & 0 & 11 \\ \hline
\end{tabular}
\caption{The number of unique property violations detected when running the testing procedure on \tcbase and \tcir for translating from Java to C++. The first column (BS) shows the beam-size parameter, and the second column the value of $k$ of the corresponding $k$-safety property. The third column shows the (abbreviated) program-transformation (PT) functions, which are combined with the (abbreviated) program-inspection (PI) functions of the first row to form the properties in\app{sect:appendix-specs}. We report the number of violations under each PI function---the left sub-column shows the violations for \tcbase and the right for \tcir. For example, the violations of the property in \Cref{fig:semantic1} are shown in the $k$=1 rows under \texttt{\textbf{retV/s}}: for BS=1, there are 254 violations for \tcbase and 234 for \tcir.}
\label{tab:java-to-cpp}
\end{table*}

\begin{table*}[ht!]
\centering
{\fontsize{9pt}{9pt}\selectfont}
\begin{tabular}{@{}|c|c|c|S[table-format=1.0]|S[table-format=3.0]|S[table-format=2.0]|S[table-format=3.0]|S[table-format=2.0]|S[table-format=4.0]|S[table-format=3.0]|S[table-format=3.0]|S[table-format=3.0]|S[table-format=3.0]|S[table-format=3.0]|S[table-format=3.0]|S[table-format=3.0]|S[table-format=3.0]|S[table-format=3.0]|@{}}
\hline
 \multirow{2}{*}{BS} & \multirow{2}{*}{$k$} & \multirow{2}{*}{\diagbox[width=1.3cm]{PT}{PI}} & \multicolumn{3}{c|}{\multirow{2}{*}{\texttt{\textbf{arity}}}} & \multicolumn{3}{c|}{\multirow{2}{*}{\texttt{\textbf{numC/s}}}} & \multicolumn{3}{c|}{\multirow{2}{*}{\texttt{\textbf{numL/s}}}} & \multicolumn{3}{c|}{\multirow{2}{*}{\texttt{\textbf{compiles}}}} & \multicolumn{3}{c|}{\multirow{2}{*}{\texttt{\textbf{retV/s}}}} \\
 & & & \multicolumn{3}{c|}{} & \multicolumn{3}{c|}{} & \multicolumn{3}{c|}{} & \multicolumn{3}{c|}{} & \multicolumn{3}{c|}{} \\ \hline
  \multirow{8}{*}{1} & 1 & -- & 0 & 0 & 1 & 8 & 2 & 14 & 9 & 17 & 13 & 103 & 84 & 32 & 211 & 237 & 294 \\ \cline{2-18}
 & \multirow{6}{*}{2} & \texttt{\textbf{rnmP}} & 0 & 0 & 0 & 36 & 32 & 77 & 46 & 23 & 961 & 188 & 177 & 101 & 131 & 147 & 437 \\
 &  & \texttt{\textbf{addP}} & 0 & 0 & 18 & 50 & 32 & 88 & 76 & 34 & 90 & 176 & 204 & 1168 & 131 & 117 & 351 \\
 &  & \texttt{\textbf{addC}} & 0 & 1 & 3 & 28 & 13 & 2162 & 58 & 36 & 78 & 131 & 216 & 299 & 626 & 428 & 206 \\
 &  & \texttt{\textbf{chBC}} & 0 & 0 & 18 & 51 & 21 & 467 & 81 & 28 & 102 & 1086 & 1184 & 466 & \si{-} & \si{-} & \si{-} \\
 &  & \texttt{\textbf{addL}} & 0 & 0 & 1 & 21 & 32 & 134 & 43 & 14 & 1053 & 209 & 208 & 286 & 621 & 650 & 629 \\
 &  & \texttt{\textbf{rmL}} & 0 & 0 & 2 & 3 & 0 & 80 & 33 & 14 & 99 & 109 & 113 & 64 & \si{-} & \si{-} & \si{-} \\ \cline{2-18}
 & 3 & \texttt{\textbf{mrg}} & 0 & 2 & 8 & 166 & 94 & 244 & 220 & 108 & 175 & 994 & 1134 & 480 & 132 & 162 & 512 \\ \cline{1-18}
 \multirow{8}{*}{3} & 1 & -- & 0 & 0 & 0 & 4 & 0 & 9 & 2 & 12 & 4 & 80 & 55 & 25 & 136 & 173 & 288 \\ \cline{2-18}
 & \multirow{6}{*}{2} & \texttt{\textbf{rnmP}} & 0 & 0 & 0 & 1 & 1 & 13 & 3 & 3 & 18 & 11 & 5 & 19 & 3 & 6 & 34 \\
 &  & \texttt{\textbf{addP}} & 0 & 0 & 11 & 0 & 2 & 7 & 5 & 5 & 19 & 12 & 10 & 77 & 7 & 1 & 54 \\
 &  & \texttt{\textbf{addC}} & 0 & 0 & 0 & 0 & 0 & 2038 & 7 & 4 & 25 & 12 & 21 & 191 & 382 & 212 & 75 \\
 &  & \texttt{\textbf{chBC}} & 0 & 0 & 0 & 16 & 8 & 279 & 7 & 5 & 26 & 706 & 878 & 382 & \si{-} & \si{-} & \si{-} \\
 &  & \texttt{\textbf{addL}} & 0 & 0 & 0 & 1 & 1 & 21 & 10 & 2 & 565 & 23 & 24 & 89 & 338 & 429 & 469 \\
 &  & \texttt{\textbf{rmL}} & 0 & 0 & 0 & 0 & 0 & 7 & 1 & 5 & 78 & 50 & 38 & 33 & \si{-} & \si{-} & \si{-} \\ \cline{2-18}
 & 3 & \texttt{\textbf{mrg}} & 0 & 0 & 0 & 19 & 11 & 82 & 30 & 16 & 62 & 368 & 546 & 1391 & 6 & 62 & 220 \\ \hline
\end{tabular}
\caption{The number of unique property violations detected when running the testing procedure on \tcbase, \dobf, and \starcoder for translating from Java to Python. Under each PI function, the left sub-column shows the violations for \tcbase, the middle for \dobf, and the right for \starcoder. For example, the violations of the property in \Cref{fig:semantic2} are shown in the $k$=2, \texttt{\textbf{rnmP}} rows under \texttt{\textbf{retV/s}}: for BS=1, there are 131 violations for \tcbase, 147 for \dobf, and 437 for \starcoder.}
\label{tab:java-to-py}
\end{table*}

\begin{table*}[ht!]
\centering
{\fontsize{9pt}{9pt}\selectfont}
\begin{tabular}{@{}|c|c|c|c|c|c|@{}}
\hline
\multirow{3}{*}{BS} & \multicolumn{5}{c|}{Average Running Time (s)} \\
& \multicolumn{1}{c|}{\tcbase} & \multicolumn{1}{c|}{\tcir} & \multicolumn{1}{c|}{\tcbase} & \multicolumn{1}{c|}{\dobf} & \multicolumn{1}{c|}{\starcoder} \\
& \multicolumn{1}{c|}{Java-C++} & \multicolumn{1}{c|}{Java-C++} & \multicolumn{1}{c|}{Java-Python} & \multicolumn{1}{c|}{Java-Python} & \multicolumn{1}{c|}{Java-Python} \\ \hline
1 & 2.0 & 2.1 & 1.5 & 1.3 & 42.9 \\ \hline
3 & 5.9 & 6.8 & 2.8 & 3.3 & 51.3 \\ \hline
\end{tabular}
\caption{The average running time (in seconds), including model inference and test harness execution, for checking a property on each model.}
\label{tab:time}
\end{table*}

\twocolumn
}
{
}
%%----------------------------------------------------------------------------
\section*{Acknowledgements}
This work was supported by DFG grant 389792660 as part of TRR 248 (see \url{https://perspicuous-computing.science}).

\bibliography{aaai24}

\end{document}